\documentclass[preprint, 10pt]{elsarticle}
\journal{Journal of \LaTeX\ Templates}

\usepackage[utf8]{inputenc}
\usepackage[margin=1in,includefoot]{geometry}
\usepackage{xcolor}
\usepackage{hyperref}
\usepackage{amsthm,amsfonts,latexsym,amssymb}
\usepackage{amsmath,amssymb}
\usepackage{bm}
\usepackage{array,multirow}
\usepackage{setspace}   
\usepackage{diagbox}
\usepackage{adjustbox}
\usepackage{comment}
\usepackage{graphicx}
\usepackage{float}
\usepackage{tabularx}
\usepackage{svg}
\usepackage{tikz}
\usetikzlibrary{positioning}
\usetikzlibrary{shapes,matrix}
\usepackage[left, modulo]{lineno}
\usepackage{multirow}

\vspace{1mm}
\begin{document}
\abovedisplayskip=6.0pt
\belowdisplayskip=6.0pt

\title{Developing a cost-effective emulator for groundwater flow modeling using deep neural operators}

\address[gk]{School of Civil Engineering, University of Leeds, Leeds, UK}
\address[my2]{School of Computing, University of Leeds, Leeds, UK}
\address[bu]{Division of Applied Mathematics, Brown University, Providence, U.S.A}
\address[my]{Deltares, Delft, The Netherlands}

\author[gk]{Maria Luisa Taccari\corref{cor1}}  
\author[my2]{He Wang } 
\author[bu]{Somdatta Goswami}
\author[my]{Jonathan Nuttall} 
\author[gk]{Xiaohui Chen}
\author[my2]{Peter K. Jimack }
\cortext[cor1]{Corresponding author. Email: cnmlt@leeds.ac.uk}

\begin{abstract} 
Current groundwater models face a significant challenge in their implementation due to heavy computational burdens. To overcome this, our work proposes a cost-effective emulator that efficiently and accurately forecasts the impact of abstraction in an aquifer. Our approach uses a deep neural operator (DeepONet) to learn operators that map between infinite-dimensional function spaces via deep neural networks. The goal is to infer the distribution of hydraulic head in a confined aquifer in the presence of a pumping well. We successfully tested the DeepONet on four problems, including two forward problems, an inverse analysis, and a nonlinear system. Additionally, we propose a novel extension of the DeepONet-based architecture to generate accurate predictions for varied hydraulic conductivity fields and pumping well locations that are unseen during training. Our emulator's predictions match the target data with excellent performance, demonstrating that the proposed model can act as an efficient and fast tool to support a range of tasks that require repetitive forward numerical simulations or inverse simulations of groundwater flow problems. Overall, our work provides a promising avenue for developing cost-effective and accurate groundwater models. 
\end{abstract}


\maketitle
\doublespacing

\section{Introduction}\label{Intro}

The computational efficiency of existing numerical models for groundwater becomes an issue when dealing with large-scale or highly nonlinear systems, particularly where decision-making relies on real-time simulation inference. The nonlinear nature of many groundwater problems necessitates the use of an iterative process to solve equations, while the computational cost can escalate rapidly when re-calibrating the initial model to incorporate new observational data \cite{farrell2017splicing}. Consequently, the heavy computational burden of groundwater models can limit their implementation in decision-making processes, as is the case for the National Water Model (NWM) used to manage drought risk in the Netherlands, for example \cite{mens2021dilemmas}.

In recent years, the use of deep learning techniques for functional approximation has gained significant attention due to their potential in developing efficient low-fidelity models that can approximate expensive numerical methods in a wide range of applications \cite{karniadakis2021physics,samaniego2020energy,kochkov2021machine,bar2019learning,jeon2021fvm}. One area where such approximations have been successful is in using deep convolutional neural networks (CNNs) as surrogates for dynamic multiphase flow problems \cite{taccari2022attention,mo2019deep,mo2019deep2,zhong2019predicting}. The deep CNN-based surrogate models treat the problem as an image-to-image regression, where the input and output functions are represented as images and the deep CNNs learn the mapping between them. The resulting models are capable of accurately predicting pressure and saturation fields with highly heterogeneous aquifer conductivity fields at arbitrary time instances. However, employing deep CNNs as surrogate models has several challenges. It is limited to problems where the input and output functions are defined on a lattice grid, and the training data encompasses all grid values within the computational domain. The solution cannot be evaluated at any arbitrary query point lying within the trained domain. The accuracy of the model and its architecture both depend on the mesh resolution, meaning that the model must be retrained for different mesh resolutions to maintain its accuracy \cite{zhu2018bayesian}. Furthermore, independent simulations need to be performed for every different domain geometry, input parameter set, or initial/boundary conditions (I/BCs). 

For the generalization of the solution, we need to look into higher levels of abstraction to learn the mapping from an input functional space to an output functional space (and not a vector space as in functional regression). To that end, the universal approximation theorem for operators \cite{Chen1995UniversalAT} is suggestive of the potential application of deep neural networks in learning nonlinear operators from data. The neural operators, introduced in $2019$ in the form of deep operator networks (DeepONet)\cite{Lu2021a}, learns the mapping between two infinite dimensional Banach spaces, providing a unique simulation framework for real-time prediction of multi-dimensional complex dynamics. Once trained, the DeepONet is discretization invariant, which means the same network parameters are shared across different parameterizations of the underlying functional data, and hence can be used to obtain the solution at any arbitrary spatial and temporal location (interpolation). Furthermore, a recent theoretical work \cite{lanthaler2022error} has shown that DeepONet can break the curse of dimensionality in the input space.

In this paper, we demonstrate through multiple problem setups that DeepONet provides fast and accurate inferences for both explicit as well as implicit operators, and hence can be employed as an efficient surrogate model in approximating various quantities of interest in the domain of subsurface flows. The readers can refer to \autoref{sec:method} for details of the method and an overview of related works. Specifically,  we employ the DeepONet framework to design an efficient emulator model to estimate the impact of abstraction in the distribution of hydraulic heads in a heterogeneous confined aquifer. We demonstrate, for the first time, that DeepONet can be applied effectively to groundwater problems and we illustrate some of the potential benefits of this learning approach in the domain of subsurface flows. The proposed method can efficiently learn solutions of both the forward and inverse problems, the latter being notoriously difficult and time-consuming with traditional methods. We successfully employ DeepONet for fast inference of a nonlinear system, which would require the use of an iterative methods using standard numerical solvers. Finally, we propose a modification to the vanilla DeepONet in order to successfully predict the distribution of spatially varying groundwater heads given a well is randomly positioned in the heterogeneous aquifer. 

The paper is organized as follows. Section~\ref{sec:problem} discusses the problem statement for groundwater flows and the four experiments analysed in this study: $i)$ mapping from the hydraulic conductivity field to the groundwater field, $ii)$ the mapping from a pumping well location and the hydraulic conductivity field to the groundwater field, $iii)$ learning an inverse mapping from the hydraulic head to the hydraulic conductivity field, and finally $iv)$ a nonlinear problem with a head-dependent boundary condition. Section~\ref{sec:method} summarizes the DeepONet approach and provides an overview of the model setup, along with training of the surrogate model. In \autoref{sec:results}, we apply the neural operator to the four experimental setups discussed above while \autoref{sec:summary} summarizes the findings and outlines the future directions.

\section{Problem statement}
\label{sec:problem}

This work focuses on learning the non-linear operator that is represented by the solution of the governing equation of groundwater flow employing a neural operator. In this section, we introduce the partial differential equation (PDE) governing the ground water flow and the computational limitations of a typical numerical solver. We then discuss four experiments which have been designed to illustrate the range of applicability of the proposed method.

\subsection{Governing Partial Differential Equation}
\label{sec:darcy}

The governing PDE that defines the movement of groundwater on a two-dimensional space combines the Darcy’s Law and the principle of conservation of mass \cite{modflow}, and is written as:
\begin{equation}\label{eq:darcy_PDE}
\begin{split}
    {S_{s}\frac{\partial h}{\partial t} -\nabla \cdot (K\nabla h) = q_{s}}, \\
\end{split}
\end{equation}
which is constrained by certain boundary conditions. In \autoref{eq:darcy_PDE}, $h$ is the hydraulic head [L], $K$ is spatially varying hydraulic conductivity field [L/T], $q_{s}$ is the volumetric flux of groundwater sources and sinks per unit volume [1/T], $S_{s}$ is the specific storage [1/L], and t [T] is time. A table describing all the notations can be found in \ref{sec:notation}.  

The U.S. Geological Survey (USGS) finite-difference flow model, MODFLOW \cite{modflow} has been broadly used for over the last $30$ years by researches, consultants, and governments to efficiently simulate groundwater flow \cite{hughes2022modflow}. The Environment Agency, which is the national environmental regulator for England and Wales, uses MODFLOW to analyze the impacts of various scenarios on the hydrological and geohydrological behavior of the principal aquifers \cite{farrell2017splicing}. The model enables the agency to investigate how the aquifer system responds to changes in water withdrawal rates, variations in recharge rates, and the introduction of new recharge or discharge sources. This information helps the agency make informed decisions about managing the country's water resources and protecting the environment. The computational time required to run a groundwater model with a spatial grid of $200$m and a temporal resolution of one to three time steps per month, for a time period ranging from months to years, can be prohibitively expensive, especially when solving optimization problems with the aim of maximizing groundwater withdrawals given some constraints. Furthermore, solving an inverse problem for inferring aquifer material properties requires multiple simulations either to discover the missing physics or to calibrate the free parameters of the formulated inverse problem. Such computational burden motivates the development of a deep neural network based emulators to provide predictions with high accuracy while substantially reducing the computational costs at run time. In this work, we have considered a deep neural operator based surrogate model that is a viable alternative to numerically approximating the governing equation for multiple input functions, and thus efficiently forecast the impact of abstraction in an aquifer.   

\subsection{Operator learning task}
\label{sec:operator_intro}

An operator, denoted by $\mathcal{G}$, is a mathematical function that takes one or more functions as input and produces another function as output. Given an input function $\bm u({\bm{x}}) \in \mathbb{R}^{d_x}$ and an output function $\bm v({\bm{x}}) \in \mathbb{R}^{d_y}$, the operator $\mathcal{G}$ is defined as $\mathcal{G} : \bm u({\bm{x}}) \in \mathbb{R}^{d_x} \mapsto \bm v({\bm{x}}) \in \mathbb{R}^{d_y}$, where $\mathbb{R}^{d_x}$ and $\mathbb{R}^{d_y}$
represents the dimensionality of the inputs and the outputs, respectively and $\bm x$ denotes the spatial and temporal coordinates which defines the output space. A PDE may be regarded as an operator: the input space consists of the functions required to specify the problem definition, such as initial and boundary conditions (ICs/BCs), forcing functions and coefficients (which may vary spatially and temporally). The output space is the Sobolev space on which the solution of the PDE lies. Our goal is to approximate the PDE introduced in \ref{sec:darcy} with a neural operator ${\mathcal{G}_{\bm{\theta}}}$ where ${\bm \theta}$ collectively represents the parameters of the neural operator, the weights, $\mathbf W$ and the biases, $\mathbf b$. The mathematical formulation of DeepONet is introduced in \autoref{sec:deeponet}. We demonstrate the effectiveness of approximating subsurface flows with a neural operator using four computational experiments which are introduced in the next section.

\subsection{Computational Experiments}
\label{sec:intro_tests}

\begin{figure}[H]
     \centering
      \vspace*{-10mm}
 \includegraphics[width=1.0\textwidth]{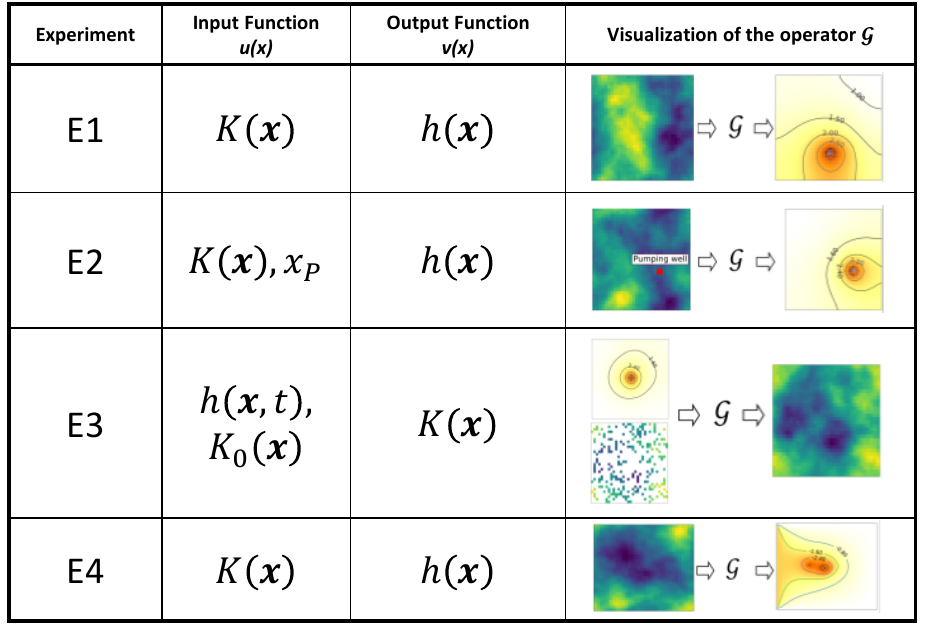}
 \vspace*{-10mm}
 \caption{A schematic representation of the experiments under consideration in this work. The input/output functions and representative plots to demonstrate the task that the operator learns are shown.}
 \label{fig:4testcases} 
 \end{figure}

The focus of this study is to develop a fast emulator for groundwater flow. We aim to demonstrate that the proposed framework can be used to eﬃciently estimate the pumping-induced change of the groundwater level, relative to the level before pumping, in a highly heterogeneous confined aquifer (Experiments $E_{1}$ and $E_{2}$). Furthermore, we employ the framework for solving inverse problems (Experiment $E_{3}$), which require a large number of forward numerical simulations if a traditional numerical solver is employed. Finally, we solve a nonlinear system (Experiment $E_{4}$), where the solution is conventionally obtained through an iterative process, and hence is heavily time-consuming. This section introduces the four computational experiments ($E_{1}$ - $E_{4}$) designed in the context of subsurface flow presented in \autoref{sec:darcy}. A visual description of different experiments considered in this work is presented in Figure \ref{fig:4testcases}. Details of the data generation to consider heterogeneity within the aquifer are presented in \ref{sec:dataset}.  

The description of the experiments, E\# are as follows.
\begin{itemize}
    \item \textbf{E1: Forward problem, $\mathcal G_{\bm \theta}: K(\bm x)\mapsto h(\bm x)$}: The goal of this experiment is to learn the solution operator, $\mathcal G_{\bm \theta}$ that maps the spatially varying conductivity field, $K(\bm x)$ to the hydraulic head, $h(\bm x)$ at some subsequent timestep. In other words, the learning goal is to infer the distribution of hydraulic head in a heterogeneous confined aquifer with one fully penetrating well that starts pumping at a constant rate $q_{s}$ at $t=0$. The solution is the prediction of the distribution of hydraulic head $h(\bm x)$ at the time instance $t=T$ given a spatially varying hydraulic conductivity field $K(\bm x)$ and under the assumption of no prescribed flows or heads along the boundary of the domain.  
    \item \textbf{E2: Multiple input functions, $\mathcal G_{\bm \theta}:[K(\bm x), x_{P}] \mapsto h(\bm x)$}: The goal of this experiment is to learn a solution operator to approximate the hydraulic head at a time $T$, given spatially varying hydraulic conductivity $K(\bm x)$ and the location of the pumping well $x_{P}$ as input functions.
    \item \textbf{E3: Inverse problem, $\mathcal G_{\bm \theta}: [h(\bm x,t),\mathcal K_{0}(\bm x)] \mapsto \mathcal K(\bm x)$}: The aim of this experiment is to learn an inverse operator that approximates the spatially varying hydraulic conductivity field $K(\bm x)$ given the hydraulic head on a domain at different time instances. Understanding that the problem does not have a non-trivial solution, we acknowledge that having more observations of the solution field increases the chances of finding a unique solution that the model converges to. To constrain the solution space in the inverse modeling process, we incorporate sparse observations of hydraulic conductivity $K_{0}(\bm x)$ as additional inputs.
    \item \textbf{E4: Nonlinear system, $\mathcal G_{\bm \theta}: K(\bm x)\mapsto h(\bm x)$:} The scenario of this experiment is that a pumping well is located in the center of the domain and a head-dependent well is fixed at a different location within the domain. While a pumping well has speciﬁed ﬂow boundaries,\textit{i.e.}, the ﬂow is not a function of the head, the speciﬁed ﬂow of the head-dependent well is calculated as a function of the hydraulic head. The goal of learning the operator $\mathcal G^{NL}$ is to approximate the mapping between the hydraulic conductivity $K(\bm x)$ of the heterogeneous aquifer and the distribution of hydraulic head $h(\bm x)$ directly. In a traditional solver, nonlinearities are resolved using an iteration loop by repeatedly formulating and solving the governing equation using heads from the previous iteration until the residual of the governing equation is within a specified tolerance. The proposed neural operator-based solution eliminates the need for iterative solvers.
\end{itemize}

\section{Solution operator approximation methods}
\label{sec:method}

In this section, we introduce the architecture of the deep operator network, DeepONet and discuss about some of the recent works where neural operators have been employed to solve PDEs. Consider two separable Banach spaces, $\bm u=\bm u(\Omega; \mathbb{R}^{d_x})$ and $\bm v=\bm v(\Omega;\mathbb{R}^{d_y})$, where $\Omega$ is a bounded open set in $\mathbb{R}^D$, and $\mathbb{R}^{d_x}$ and $\mathbb{R}^{d_y}$ are the dimensionality of the inputs and the outputs, respectively. The nonlinear map $\mathcal{G}$, arising from the solution of a time-dependent PDE (\autoref{eq:darcy_PDE}) at some time $T$, maps from $\bm u$ to $\bm v$. The objective is to approximate the nonlinear operator $\mathcal{G}$ through a parametric mapping as $\mathcal{G}: \bm u \times \bm \Theta \rightarrow \bm v$ or $\mathcal{G}_{\bm \theta}: \bm u \rightarrow \bm v$, where $\mathcal{G}_{\bm \theta}$ represents the parametric mapping and $\bm \Theta$ is a finite-dimensional parameter space. The optimal parameters $\bm\theta^*$ are found by training a neural operator using backpropagation on a dataset of $\{\mathbf{u}_j, \mathbf{v}_j \}_{j=1}^N$ generated on a discretized domain. 

\subsection{Deep Operator Network}
\label{sec:deeponet}

\begin{figure}[H]
    \centering
    \includegraphics[width=0.8\textwidth]{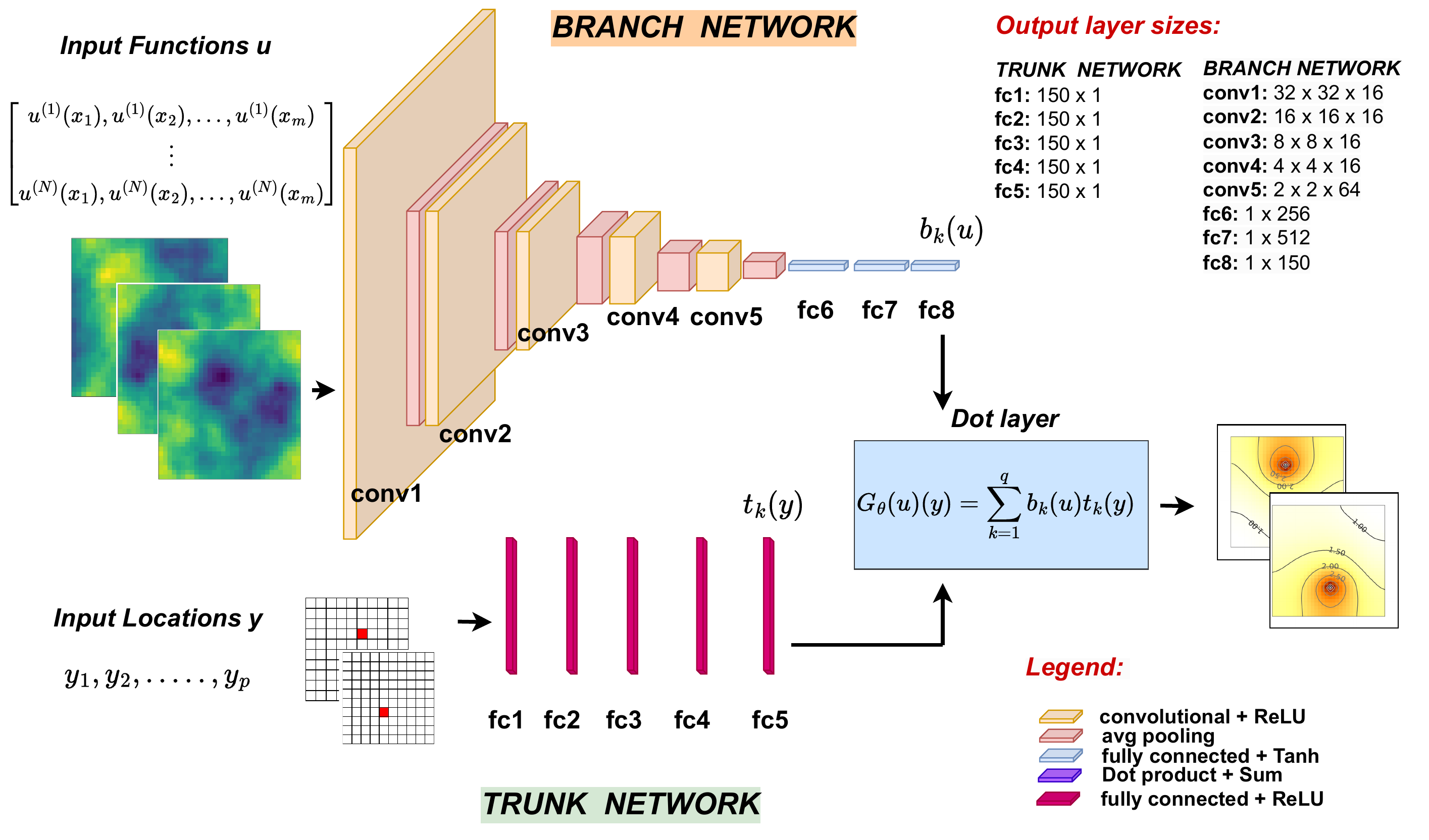}
    \caption{Schematic representation of the network architecture of vanilla DeepONet employed in this work. In this work, we have considered a CNN as a branch net and a fully connected feed forward neural network as trunk net. The outputs of the branch and the trunk networks are combined through an inner product to approximate the solution operator.}
    \label{fig:diagram_vanillaDeepONet}
\end{figure}

The universal approximation theorem for operators by proposed by Chen and Chen \cite{Chen1995UniversalAT} states that shallow neural networks, of sufficient width, are capable of approximating any nonlinear continuous functional or operator to arbitrary accuracy. This theorem is based on a particular neural network model which is composed of two concurrent sub-networks and the outputs of the networks are combined by an inner product. Motivated by the universal approximation theorem, the deep operator Network (DeepONet) \cite{Lu2021a} was proposed to learn the mapping between Banach spaces with infinite dimensions. The DeepONet architecture consists of two deep neural networks (DNNs): the branch net encodes the input function, $\bm u$, at fixed sensor points, $\{x_1, x_2, \dots, x_m\}$, while the trunk net encodes the information related to the spatio-temporal coordinates, $\zeta = \{x_i, y_i, t_i\}$, at which the solution operator is evaluated to compute the loss function. The learning process takes place in a general setting, meaning that the sensor locations (${x_i}_{i=1}^m$) at which the input functions, $\bm u$ are evaluated don't have to be evenly spaced, but they must be consistent across all input function evaluations. The branch net takes $[\bm{u}(x_1), \bm{u}(x_2), \dots, \bm{u}(x_m)]^T$ as input and outputs $[b_1, b_2, \ldots, b_q]^T \in \mathbb{R}^q$, while the trunk network takes $\zeta$ as input and produces $[t_1, t_2, \ldots, t_q]^T \in \mathbb{R}^q$ as output. These two subnetwork outputs are combined through a dot product to produce the desired result. A bias ($b_0 \in \mathbb{R}$) is added in the final stage to increase expressiveness, resulting in $\mathcal{G}(\bm{u})(\zeta) \approx \sum_{i=k}^q b_k t_k + b_0$. The optimized values of the trainable parameters $\bm{\theta}$ can be obtained by minimizing a mean square error loss function. Figure \ref{fig:diagram_vanillaDeepONet} illustrates the architecture of the vanilla DeepONet proposed in \cite{Lu2021a}. 

\subsection{Related Works}
\label{sec:related_work}

DeepONet has shown remarkable success in diverse fields of applications like approximating irregular ocean waves \cite{cao2023deep}, learning stiff chemical kinetics \cite{goswami2023learning}, bubble dynamics \cite{lin2021operator}, microstructure evolution \cite{oommen2022learning} \textit{etc.}, where the network is trained using large datasets for solving a forward mode problem. Additionally, some recent work has been focused on learning the mapping of multiple input functions to the solution field \cite{jin2022mionet,goswami2022neural}. Prior work of DeepONet in the area of subsurface flow problems has been to learn the mapping from the conductivity ﬁeld to the hydraulic head in simple and complex geometries through data-driven \cite{FNOvsDON} and physics informed approaches \cite{goswami2021physics,goswami2022physics_fracture,wang2021learning}. In both these works, the PDE governing the subsurface flow (Darcy's equation) has been employed as an application to demonstrate the framework proposed in the corresponding work. In another study, an operator level transfer learning framework \cite{TL_conditional} was proposed, where Darcy's equation was employed as an example to demonstrate the approach. The idea behind operator transfer learning is to train a source model with sufficient labeled data from a source domain under a standard regression loss and transfer the learned variables to a second target model, which is trained with very limited labeled data from a target (different but related) domain under a hybrid loss function that is the sum of the regression loss and a conditional embedding operator discrepancy loss. Furthermore, another operator-level transfer learning framework was proposed in \cite{kahana2023geometry}, where Darcy's equation was solved on an L-shaped domain (source) and transferred to an L-shaped domain with a hole. The implementation of a hybrid solver (HINTS) approach could directly handle the change in target geometry and does not require retraining of the operator. None of the works discussed above deal with the additional specific storage term in \autoref{eq:darcy_PDE} and is not dedicated to employing the operator for multiple scenarios in subsurface flows. Furthermore, for the first time, the operator network is designed to solve an inverse problem in \textbf{E3} to learn the hydraulic conductivity field, $K(\boldsymbol x)$, from the hydraulic head, $h(\boldsymbol x)$ and sparse observations of $K(\boldsymbol x)$.

In an independent work, Lanthaler et al. provide a theoretical analysis of the approximation and generalization errors of DeepONet \cite{lanthaler2022error}. They accomplish this by decomposing the error into encoding, approximation, and reconstruction errors and theorizing the lower and upper bounds of the total error. Their analysis indicates that the accuracy of DeepONet can deteriorate in the presence of discontinuities or sharp features that are not fixed in space, while DeepONet can accurately learn operators that produce discontinuities or sharp features fixed in space. This is in-line with our observation and we propose a modified architecture to deal with such scenarios. According to Hadorn \cite{Shift-DeepOnet}, it struggles to learn sharp features for each location without increasing the basis functions from the trunk net. Unfortunately, increasing basis functions becomes infeasible for high dimensional problems. An effective modiﬁcation to overcome this bottleneck of DeepONet to deal with translational invariance is to eliminate the invariance. Both Shift-DeepONet \cite{Shift-DeepOnet} and FlexDeepONet \cite{Flex-DeepONet} add pre-transformation sub-networks to shift, rotate and scale the data. The input functions to the branch network are passed through these additional networks, which learns the re-scaling and re-centering of these functions. A transformation layer combines the learnt shift, rotation and scale parameters with the spatial coordinates of the evaluation points: the outputs of this layer are the inputs of the trunk network, such that the basis functions of the trunk net depend on the input functions. Similarly, another extension of DeepONet introduces two encoders one each to the inputs of the branch and the trunk network \cite{wang2021fast}. The embedded features are inserted into the hidden layer of both sub-networks using point-wise multiplication. This novel architecture appears to be more resilient than the conventional DeepONet architecture to vanishing gradient pathologies. 

In the current work, we consider the vanilla version of the DeepONet as ﬁrstly introduced in \cite{Lu2021a} which has the benefit of a simpler architecture. In the later part of the work, we propose a modified version of DeepONet in order to overcome the limitations of the vanilla DeepONet in dealing with a source term which is not always defined at the same location, and leads to sharp gradients in the solution field. The next section presents the details of the employed network architectures.

\subsection{Network architecture and training} 
\label{architecture}

The branch net is considered as a convolution neural network (CNN) that takes as input the functions, $\bm{u}$ evaluated on a lattice grid of size $32 \times 32$, which is consistent for all the experiments carried out in this work. For experiments \textbf{E1} and \textbf{E4}, we have a the CNN with one input channel, however, for \textbf{E2} and \textbf{E3} we have two input channels, where the second channel denotes the location of the well and sparse observations of the target hydraulic conductivity, respectively.
The inputs to the trunk net are the coordinate of $128$ evaluation points, which are randomly sampled in the domain and are distinct for each training sample. The details of the network architecture for all the experiments is provided in Table \ref{tab:architecture_detail}. A schematic representation of the network is shown in Figure \ref{fig:diagram_vanillaDeepONet} 

\begin{singlespace}
\begin{table}[!htb]\small 
\caption{Architecture details of vanilla DeepONet employed for all the experiments (\textbf{E1}-\textbf{E4}).}
\centering
\begin{tabular}{llcccccc}
    \hline
     & \textbf{Layer} & \textbf{Kernel Size} & \textbf{Width} & \textbf{Activation} & \textbf{Output}\\ \hline
     \multicolumn{6}{ c }{Branch Network} \\ \hline
     1 & Conv2D & $5\times5$ & $16$ & ReLU & $32\times32\times16$\\
     2 & Avg-Pool & $2\times2$ &  &  & $16\times16\times16$\\
     3 & Conv2D & $5\times5$ & $8$ & ReLU & $16\times16\times16$\\
     4 & Avg-Pool & $2\times2$ &  &  & $8\times8\times16$\\
     5 & Conv2D & $5\times5$ & $4$ & ReLU & $8\times8\times16$\\
     6 & Avg-Pool & $2\times2$ &  &  & $4\times4\times16$\\
     7 & Conv2D & $5\times5$ & $4$ & ReLU & $4\times4\times16$\\
     8 & Avg-Pool & $2\times2$ &  &  & $2\times2\times16$\\
     9 & Conv2D & $5\times5$ & $4$ & ReLU & $2\times2\times64$\\
     10 & Avg-Pool & $2\times2$ &  &  & Reshaped to $1\times64$\\
     11 & Fully connected & & $1024$ & Tanh & $1\times256$\\ 
     12 & Fully connected & & $1024$ & Tanh & $1\times512$\\ 
     13 & Fully connected & & $1024$ &  & $1\times150$\\ \hline
     \multicolumn{6}{ c }{Trunk Network} \\ \hline
     14 & Fully connected & & $150$ & ReLU & $150\times1$\\ 
     15 & Fully connected & & $150$ & ReLU & $150\times1$\\ 
     16 & Fully connected & & $150$ & ReLU & $150\times1$\\ 
     17 & Fully connected & & $150$ & ReLU & $150\times1$\\ 
     18 & Fully connected & & $150$ & ReLU & $150\times1$\\ \hline
    \end{tabular}    
    \label{tab:architecture_detail}
\end{table}
\end{singlespace}

\begin{figure}[H]
  \centering
    \includegraphics[width=0.8\textwidth]{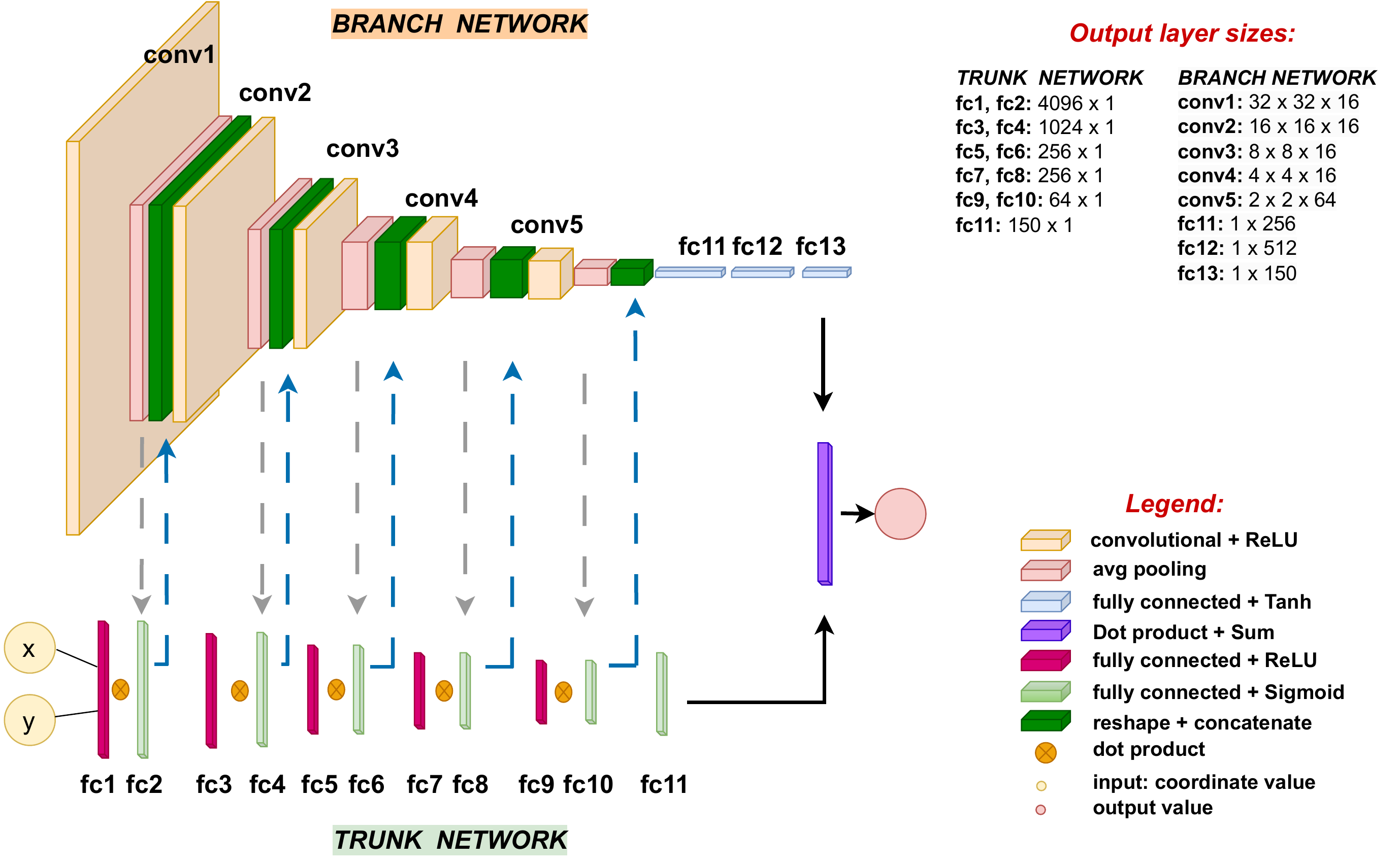}
    \caption{Schematic of the novel DeepONet architecture proposed for experiment \textbf{E2}. The architecture is specifically designed to take into account the varying locations of the pump. The basis functions approximated using the trunk net can be modified according to the position of the well.}
        \label{fig:novel_arch}
\end{figure}

In the vanilla DeepONet architecture, the solution operator is approximated as the sum over the products of the outputs of the branch and the trunk net. However, for experiment $E2$, we noticed that informing the trunk network about the location of the pumping well (input function) is key for good learning. For this reason, we propose a novel DeepONet architecture. As illustrated in Figure \ref{fig:novel_arch}, each output of the pooling layers of the branch network is combined with the output of each layer of the trunk net. The tensor coming from the branch net is flattened and followed by a dense NN layer with \textit{Sigmoid} activation function. Given the fact that the resulting vector (whose weights can be interpreted as coeﬃcients) has the same dimension of the corresponding hidden layer of the trunk net, the two vectors can be merged via an inner product. The result propagates through the following layers of both the trunk net and, after being reshaped and concatenated to the output of the pooling layer, the branch net. We demonstrate that this architecture can accurately predict the high gradients of the hydraulic head in the experiment \textbf{E2}, for which the vanilla DeepONet gives a smoother prediction. 

The implementation is carried out using the JAX framework on a single NVIDIA GeForce RTX $3080$. For all test cases, the datasets consist of $N_{train} = 1000$ training data and $N_{test} = 200$ test data. The network is trained using the Adam optimizer \cite{adam} with an initial learning rate of $5 \times 10^{-4}$ which exponentially decays every $1000$ iterations with a rate of $0.9$ and a batch size of $100$ for maximum $10^5$ iterations. We monitor the loss after every $100$ iterations and trigger an early stopping if the value of the loss for the test data does not decrease after $2 \times10^4$ iterations. 

\section{Results}
\label{sec:results}

In this section, we demonstrate through the previously discussed experiments that DeepONet can be employed for approximating a range of groundwater flow simulation problems, accurately and efficiently. We also provide a comparative study of the two architectures: the vanilla DeepONet and the novel DeepONet architecture proposed in this paper. All models are trained on a few sparse points defined on the domain but are evaluated over the whole domain for the test data. The mean relative error (MSE) is used as an error metric which corresponds to the loss function used during training, calculated as the square difference between the target and the predicted fields for all the models and all the experiments. In \ref{comparison}, we provide a comparative study of the accuracy of DeepONet with two other popular deep neural network architectures for experiments \textbf{E3} and \textbf{E4}.

\subsection{$E1$: Forward problem for fixed well location}
\label{subsec:forward_problem}

\begin{figure}[H]
    \centering
    \hspace*{-10mm}
    \includegraphics[trim={2cm 0 0 0},clip, width=1.0\textwidth]{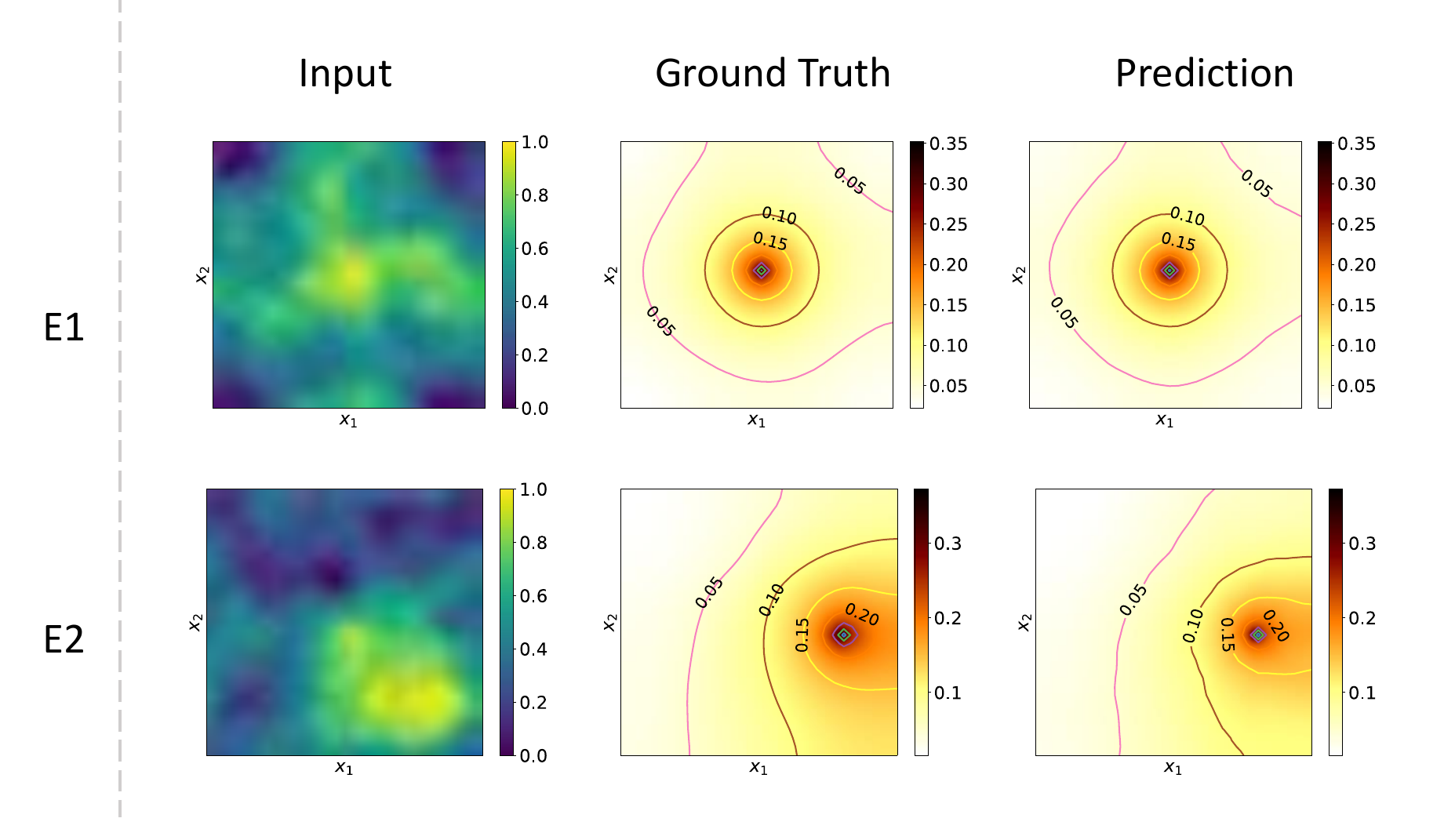}
    \vspace*{-10mm}
    \caption{Top row: Prediction of the hydraulic head, $h(\boldsymbol x)$ obtained from a trained vanilla DeepONet (Experiment $E1$) for a test sample with an unseen heterogeneous hydraulic conductivity field. The location of the well fixed at $(x,y) = (16, 16)$, which is the same for the training and the testing dataset. Bottom Row: Prediction of the hydraulic head obtained using the vanilla DeepONet where the test sample comprises unseen heterogeneous hydraulic conductivity fields and unseen pumping well location (Experiment $E2$). The results shown in this plot are for the case where the input to the branch network is the hydraulic conductivity field (first column) and the input to the trunk network is the coordinates in which the network evaluates the solution and also the well location coordinates. The second and the third column denote the ground truth and the prediction of DeepONet for the hydraulic head, respectively.}
    \label{fig:results_F1}
\end{figure}

In this experiment, the location of the well is considered the same for the training and the testing dataset and the operator learns the mapping: $\mathcal G_{\bm \theta}^{F1}: K(\bm x)\mapsto h(\bm x)$.
Training the network takes $283$ second for a total of $24\small{,}600$ iterations, and the error metrics are computed as $MSE_{train} = 1.8 \times 10^{-5}$ on $N_{train}$ samples and $MSE_{test} = 2.5 \times 10^{-4}$ on $N_{test}$ samples.
Figure \ref{fig:results_F1} (top row) shows a typical comparison between the predicted and target values of the hydraulic head given a heterogeneous hydraulic conductivity field. Both the inputs and the outputs are normalized along each individual channel, that is the variables are re-scaled into the range $[0,1]$. As can be seen in the prediction plot, DeepONet can predict the pressure buildup and the sharp increase of hydraulic head around the well very accurately. We conducted multiple analyses by altering the number of neurons, layers, kernel size, batch-size, and training data, among other factors. We observed only slight changes in the error rate, and significant deterioration occurred only when the data was not normalized or the learning rate was high. Additionally, increasing the number of query points or sampling more frequently around the well did not improve the accuracy of predictions.  

\subsection{$E2$: Forward problem for varying well locations} 
\label{results_well_loc}

\begin{figure}[H]
    \centering
    \includegraphics[width=\textwidth]{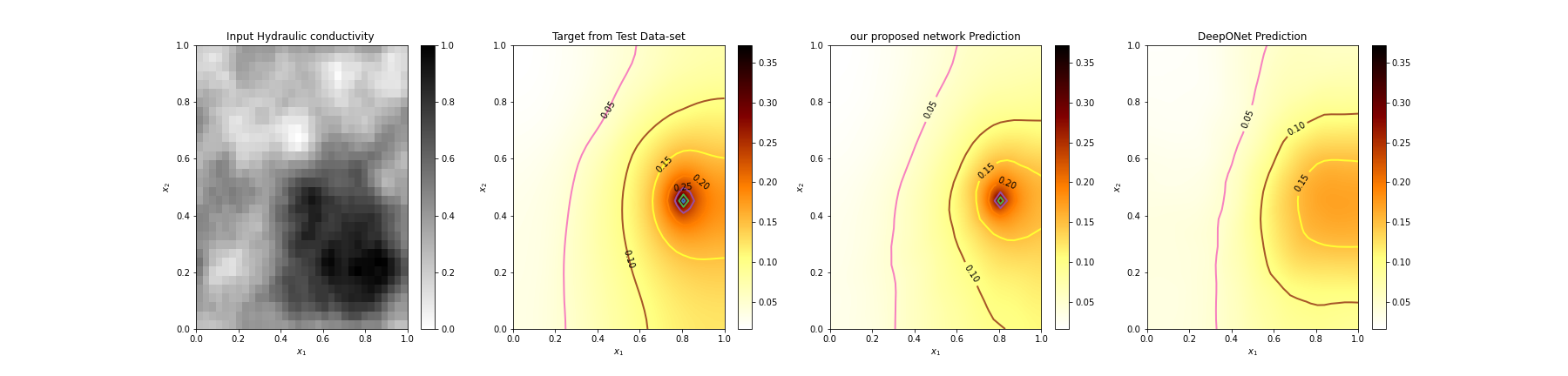}
    \includegraphics[width=\textwidth]{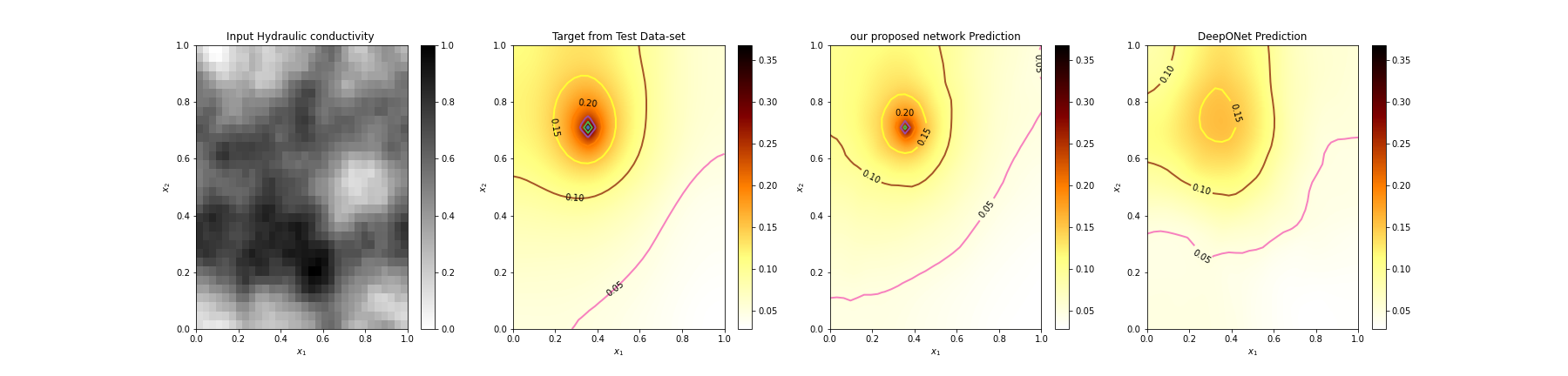}
     \includegraphics[width=\textwidth]{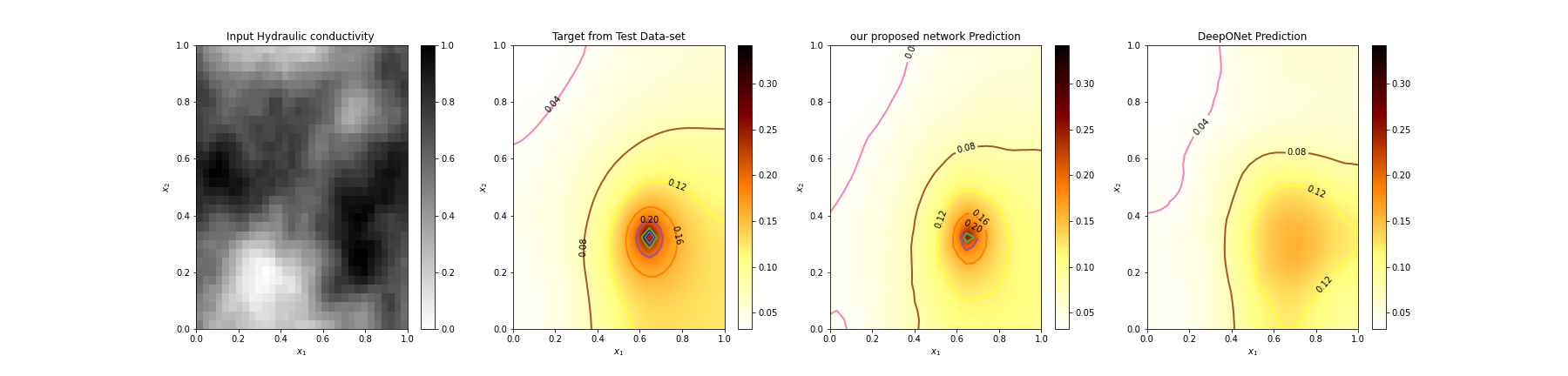}
    \vspace*{-10mm}\caption{Predictions of the hydraulic head for unseen heterogeneous hydraulic conductivity fields (first column) and unseen location of the source term (the well). Predictions with vanilla DeepONet (last column) and proposed network architecture (third column) is compared with the target fields (second column) for three representative test samples. Input to the branch network is the hydraulic conductivity field and a binary map indicating the location of the well. Input to the trunk network is the coordinates in which the network evaluates the solution.
    } 
    \label{fig:results_F2_proposed}
\end{figure}

In the next experiment, we  further expand the network capabilities to learn the solution for unseen locations of the pumping well. According to the original formulation of the DeepONet, the branch network encodes the input functions, which are the hydraulic conductivity field and the position of the source term (the well). We observed that the vanilla architecture could not capture the sharp gradients located in the region near the pumping well when the well was shifted to several locations. The reader can refer to the last column of Figure \ref{fig:results_F2_proposed} for visualization of vanilla DeepONet predictions for three representative test samples, where the location of the well (forcing term) is encoded as a binary map and concatenated to the hydraulic conductivity field as an additional channel of the existing CNN of the branch net. Through a set of computational experiments, we explored different possible ways to encode the forcing term as an input to the branch net. Those included giving the location of the source term as the set of coordinates in the Cartesian coordinate system or as the distance (magnitude and angle) between each point of the domain and the location of the well or as a Gaussian function centered at the location of the well. Within this context, we modified the network architecture by using either a single network, employing two separate neural networks for each input function, and also using two parallel networks with connections between them. All these experiments lead to either very inaccurate predictions with a wrong determination of the extent and location of the pressure front, or to smoother predictions near the source terms as previously highlighted in the last column of Figure \ref{fig:results_F2_proposed}. \ref{SVD} explores the reason for the lower prediction accuracy of the vanilla DeepONet for these test cases through the lens of the singular value decomposition (SVD). 

Finally, we found that informing the trunk net of the location of the source term is necessary for good learning in this class of problem. Figure \ref{fig:results_F1} (bottom row) shows the network prediction for one representative test case when the input to the trunk network is the concatenation of the coordinates in which the network evaluates the solution and the coordinates of the pumping well. Input to the branch network is the hydraulic conductivity field evaluated in the whole domain. Visual inspection of the results reveals that the predictions match the reference solutions very well for different distributions of $K(\boldsymbol x)$ and for varying locations of the pumping well. 
The error metric computed on the training and the testing dataset is computed to be $2.4\times10^{-5}$ and $2.7\times10^{-4}$, respectively. This approach however becomes inefficient in the case of more complex scenarios, such as multiple wells within the domain, different pumping rates of the wells, or features which are not localized at a single point (rivers and drains). Similarly to real applications, for which the modeler has a planned view of the groundwater system, we decide to encode the location of the source term as a binary map, which is concatenated to the hydraulic conductivity field as input to the branch net. As our experiments showed that informing the trunk network with the location of the forcing term is key for good learning when the location of the forcing term varies among the training data, we link the branch and the trunk net with the newly proposed architecture of DeepONet (Figure \ref{fig:novel_arch}) described in \ref{architecture}. As Figure \ref{fig:results_F2_proposed} shows for three representative test samples, the architecture that links the hidden layers of the two sub-networks significantly outperforms the vanilla DeepONet. Training takes $324$ seconds and the resulting error metrics is computed to be equal to $6.6\times10^{-5}$ and $2.6\times10^{-4}$ on the training and testing dataset, respectively. It is interesting to note that the order of magnitude of the training and testing error is always the same for the cases of the forward problems, for which the visual results are also highly satisfactory. It is reasonable to conclude that the error obtained corresponds to the lowest bound of DeepONet for the given forward problem.   

\subsection{$E3$: Inverse problem}

\begin{figure}[H]
    \centering
    \includegraphics[trim={2cm 0 0 0},clip, width=\textwidth]{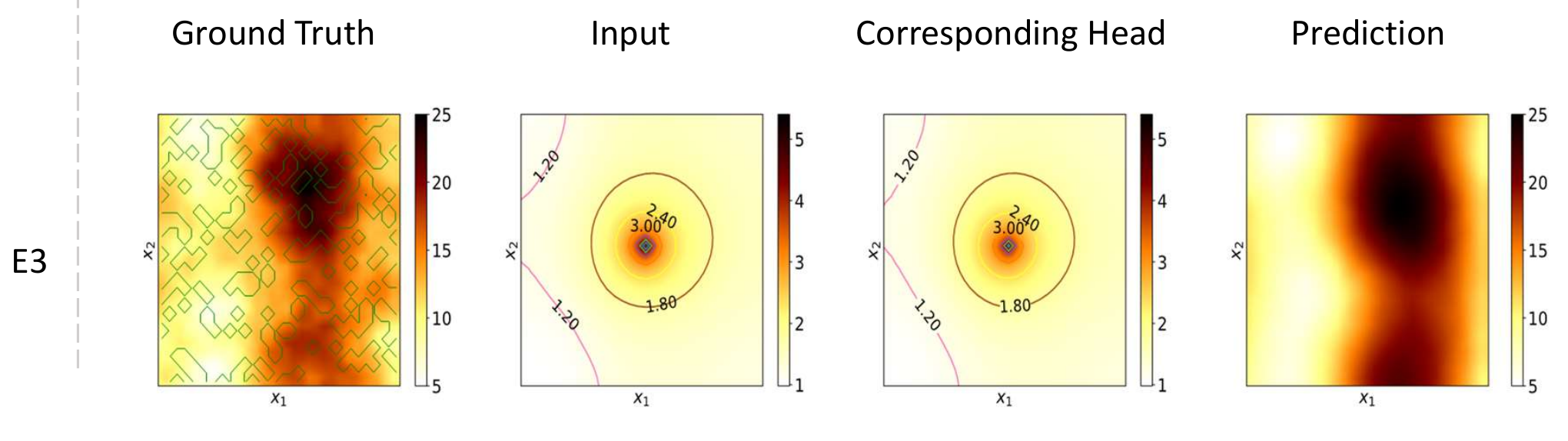}
    \vspace*{-10mm}\caption{Representative text sample for the inverse analysis using vanilla DeepONet. The first column shows the unknown hydraulic conductivity field with the location of the observation points on top of it, which are encoded as extra input in addition to the hydraulic head in the whole domain (second column). The last column shows the simulated inverse results and the third column is the hydraulic head corresponding to the predicted conductivity as calculated with the traditional solver.}
    \label{fig:results_inverse}
\end{figure}

In this section, we explore the efficiency of the operator network for solving inverse problems. More specifically, we aim to assist with underground property characterization. Given that it is impossible to directly observe the whole underground system, the aim of the inverse analysis is to understand the heterogeneous aquifer properties (\textit{i.e.}, hydraulic conductivity field) using sparse observations of the conductivity field and some known information of the hydraulic head. We consider a test problem for which the information on the hydraulic head in the whole domain is available along with sparse observations of hydraulic conductivity. In Figure ~\ref{fig:results_inverse}, we show a representative test case with the inputs and the prediction obtained from the operator network. The error metrics computed on the test samples are obtained as $1.21\times10^{-1}$ when using $20\%$ of the values of the hydraulic conductivity field (randomly sampled) as observation points. Beyond the comparison between the target reference fields (first column) and the simulated inverse results (fourth column), we also compare the input hydraulic head (second column) and the hydraulic head corresponding to the predicted conductivity as calculated with the traditional solver (third column). The operator network gives trustworthy predictions with accurate and consistent performance across the whole test dataset. The use of observational data informs the network and enhances its accuracy by $16\%$ when compared to the case in which no measurements of hydraulic conductivity were made available. An additional improvement can be achieved by informing the network with the observation of the hydraulic heads at different timesteps. When the branch net additionally encodes the hydraulic head at three equally spaced time instances (instead of only at a single time step), the predictions have a further improvement, which on average is $4\%$ more accurate. In the future, this approach could be further developed to guide efficient observational campaigns which would enhance the accuracy of the inverse models while minimizing the need for observations.

\subsection{$E4$: Nonlinear case} 
\label{results_nonlinear}

Many groundwater processes exhibit nonlinear behavior and hence requires the use of iterative solvers to obtain the solution of the hydraulic head. In this study, we show that the proposed data-driven method has the capability to resolve the system directly and save a lot on the computational cost required on the conventional iterative solver. The traditional ﬁnite-difference form of the groundwater flow equations can be written as: $\bm{A} \bm{h} =\bm{b}$, where $\bm{h}$ is the vector of head values at the end of time step, $\bm{A} $ is the matrix of the coefﬁcients of head and $\bm{b}$ is a vector of the constant terms \cite{modflow}. In a nonlinear system, the individual entries in $\bm{A}$ matrix is a function of the hydraulic head and the system of equations needs to be resolved through a nonlinear outer iteration loop. In our example, a pumping well is located in the center of the domain and a head-dependent well is fixed in another location in the domain (see \ref{sec:dataset} for full details of the nonlinearity). The hydraulic head predicted by the neural operator perfectly matches the target values in the whole domain for given hydraulic conductivity fields (considered as input to the branch net), unseen during training (Figure \ref{results_nonlinear}). The training process takes $430$ seconds for a total of $44100$ iterations and the error metrics computed on the training and the testing datasets are equal to $3.1\times10^{-5}$ and $3.9\times10^{-5}$, respectively. 

\begin{figure}[H]
    \centering
    \includegraphics[trim={2cm 0 0 0},clip, width=\textwidth]{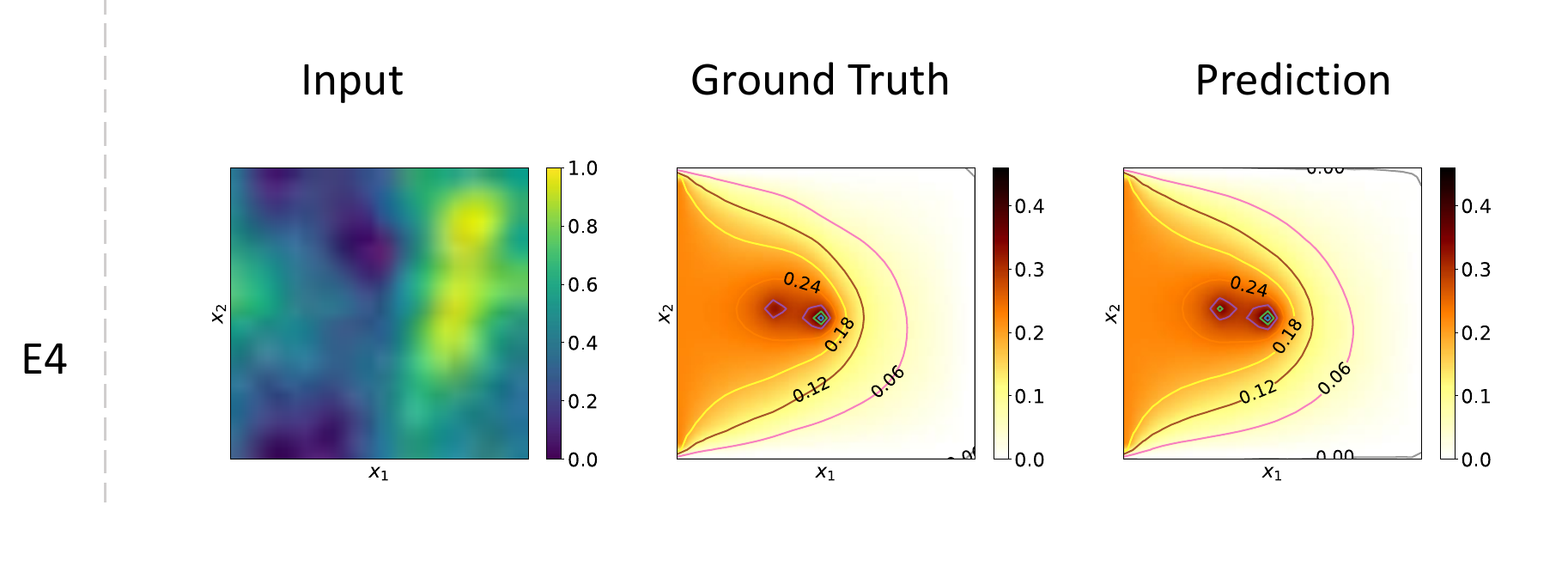}
    \vspace*{-10mm}\caption{Comparison between the ground truth and the prediction of the hydraulic head from DeepONet for unseen heterogeneous conductivity field for the nonlinear problem, \textbf{E4}.}
    \label{fig:case3_35wells}
\end{figure}

\section{Summary and Discussion}\label{sec:summary}

This paper presents the DeepONet framework as a surrogate model to efficiently and accurately calculate the state response of a groundwater system. The model is trained and tested in four experiments that demonstrate its capability to predict hydraulic head in a heterogeneous aquifer with a fixed pumping well, generalize to unseen pumping well locations, characterize aquifer properties (inverse analysis), and deal with nonlinear systems. The proposed model accurately learns both the forward and inverse relations between the spatially varying hydraulic conductivity and the hydraulic head fields very accurately. However, modifications to the original formulation of DeepONet are needed when the pumping well is placed at any location in the domain. To address this, the paper introduces a novel contribution by linking the input of the branch network to the trunk network, allowing the network to accurately predict solutions for unseen well locations and hydraulic conductivity fields. By successfully implementing the neural operator on several examples, we demonstrate the capacity of the network to support a range of tasks that require repetitive forward numerical simulations of the groundwater model. 

In the future, such a model will be extended to accommodate more complicated and realistic sub-surface problems. These could include more complex predictions from a wider range of abstraction rates and aquifer system geometry, properties, and boundaries, and the interaction with other surface water abstractions and discharges. 

\section*{Authors’ contribution}
\textbf{Maria Luisa Taccari}: conceptualization, methodology, software, validation, formal analysis, data curation, writing – original draft, writing – review \& editing, visualization. 
\textbf{He Wang}: methodology, formal analysis, writing – review \& editing, supervision.
\textbf{Somdatta Goswami}: methodology, writing – review \& editing, supervision, visualization. 
\textbf{Jonathan Nuttall}: conceptualization, methodology, software. 
\textbf{Chen Xiaohui}: supervision, funding acquisition.
\textbf{Peter K. Jimack}: methodology, formal analysis, writing – review \& editing, supervision.

\section*{Code and data availability}
All codes used to generate the datasets and train the model will be made available at  \url{https://github.com/mlttac/DeepOnet_gwf}.

\section*{Acknowledgments}
This work was supported by the Leeds-York-Hull Natural Environment Research Council (NERC) Doctoral Training Partnership (DTP) Panorama under grant NE/S007458/1. We would like to acknowledge the support provided by Deltares and we express our sincere gratitude to Bennie Minnema for his invaluable contribution to designing the nonlinear test case.

\bibliographystyle{elsarticle-num} 
\bibliography{references}

\newpage
\appendix

\section{Table of notations} 
\label{sec:notation}
A table of notations is given in Table~\ref{tab:notation_table}.

\begin{table}[H]
\centering
\begin{tabular}{c l}
 \hline
Notation & Meaning    \\
 \hline
$h$ &  Hydraulic head \\
$K$ &  Hydraulic conductivity \\
$q_{s}$ &  Volumetric flux of groundwater sources and sinks per unit volume \\
$S_{s}$ &  Specific storage \\
t &  Time  \\
$x_{P}$ & Location of the pumping well \\
$\mathcal G$ & Nonlinear Operator \\
$\mathcal G_{\bm\theta}$ & Network operator approximator\\
$\bm \theta$ & Learnable parameters\\
$u$ & Input function \\
$u$ & Input function \\
$\bm{y_j}$ & Evaluation point\\
m & Number of sensor points\\
P & Number of evaluation points\\
$N$ & Number of input functions\\
${N}_{T}$ & Number of training sample pairs\\
$q$ & Number of tunable weights at the last hidden layers of the two sub-networks \\
$\bm{b}$ & The output vector defined by the last layer of the branch network\\
$\bm{t}$ & The output vector defined by the last layer of the trunk network\\
 \hline
\end{tabular}
\caption{Table of notations}
\label{tab:notation_table}
\end{table}

\section{Numerical simulation settings for generating the datasets}\label{sec:dataset}

The labeled high-ﬁdelity datasets for constructing the data-driven (MSE) loss are simulated by solving the governing equation using a finite difference scheme and the parameters and settings are described in this section. The datasets are generated within the Python ecosystem and, in particular, using a script adapted from \cite{Olsthoorn} for the first three test cases and the iMOD Python package for the nonlinear case \cite{imodpython}. 

\begin{description}

\item[Forward problem]
The problem consists of a single-layer model representing a confined aquifer. The layer is $50$ m thick and the extent of the model R is $8 \times 10^3$ m. The grid is defined as $32\times32$. Each cell is $250$ m on a side, which corresponds to a typical spacial grid used in practice \cite{farrell2017splicing}. The aquifer, whose specific elastic storage  $S_{s}$ is $ 2 \times 10^-4$ 1/m, is highly heterogeneous with horizontal hydraulic conductivity $k \in[5, 25]$ m/d. We create the hydraulic conductivity distributions using Gaussian random fields with Power Spectrum $P(l)=l^{q}$ with $q=-4$. The resulting continuous float values are re-scaled in the range $k \in[k_{min}, k_{max}]$. The minimum and maximum values of $K$, $k_{min}$ and $k_{max}$, vary for each sample: they are randomly selected integers so that $k_{min}\in[5, 10]$ m/d and $k_{max}\in[20, 25]$ m/d. A pumping well, placed at $(x,y) = (16, 16)$, is extracting water from the sandy aquifer. Abstraction is specified at a constant rate of $Q = 5000$ m$^3$/d, which  corresponds to a typical abstraction value \cite{farrell2017splicing}. This generates the target hydraulic head, which is the solution at the time step corresponding to the simulation time $T = 200$ days. 

\item[Multiple input functions]
In the first test case, the source term is fixed in the same location and the operator learns the mapping from the hydraulic conductivity field $K$ to the groundwater field, $h$. In the second test case, the pumping well is randomly placed in the domain and the operator learns the mapping from both the well location, $x_P$ and the $K$ field to the output function. All other settings are kept the same as discussed above.

\item[Inverse problem]
The learning goal of this problem is to infer the hydraulic conductivity in the whole domain given the distribution of hydraulic head resulting from water abstraction. The numerical settings used for the inverse problem are the same as the forward problem, with the only difference being that the pumping well is now located in the center of the domain. The input and target data of the forward problem become the target and the input for the inverse problem. In addition to the hydraulic head calculated from the simulation at time $T$, we also provide sparse observations of the target hydraulic conductivity as an additional input. 

\item[Nonlinear system]
The problem consists of a single-layer model representing a confined aquifer. The layer is $60$ m thick with a surface level of $0$ m and the extent of the model R is $3.2 \times 10^4$m. The domain is discretized as $32\times32$ and each cell is $1000$ m on a side. The specific storage of the aquifer $S_{s} = 1.6 \times 10^{-4}$ [1/m] and its hydraulic conductivity varies along the domain and for each data sample. The K fields of the heterogeneous aquifer are Gaussian random fields with Power Spectrum $P(l)=l^{q}$ with $q=-4$, with values re-scaled in the range $k \in[k_{min}, k_{max}]$, where $k_{min} = 5 \times 10^-4 m/d$ and $k_{max} = 5 \times 10^-3 m/d$. Figure \ref{fig:testcase_nonlinear_setup} illustrates the set-up of the model. The initial head corresponds to the surface level everywhere. We impose Dirichlet boundary conditions with constant-head cells equal to $-2$ m along the boundary x=0 and  equal to the surface levels along the other three sides of the domain. A pumping well is located in the center of the domain,\textit{i.e.}, at  $(x,y) = (16, 16)$. The pumping well has speciﬁed ﬂow boundaries: the ﬂow is constant in time and equal to $0.5$ m$^2$/d. A head-dependent well, whose ﬂow is calculated as a function of the head, is located at $(x,y) = (10, 16)$. The relationship between the ﬂow and head of this well is reported in Figure \ref{fig:testcase_nonlinear_setup} - Right. 

\begin{figure}[H]
    \centering
    \includegraphics[width=0.4\textwidth]{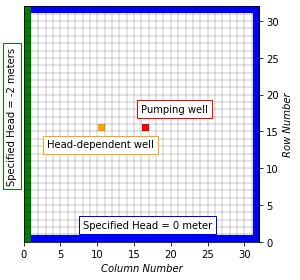}%
    \includegraphics[width=0.4\textwidth]{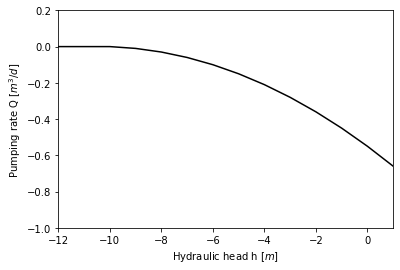}
    \caption{Setup for the inverse problem test case. Left: Boundary conditions and Source terms; Right: function between ﬂow and head for the head-dependent well.}
    \label{fig:testcase_nonlinear_setup}
\end{figure}

A simple way to specify this user-defined function is by superposing boundary conditions that are already implemented in MODFLOW. A drain is a type of boundary condition that removes water from the aquifer at a rate, called conductance, proportional to the difference between the head in the aquifer and the drain elevation. If the head in the aquifer is below the drain elevation, the drain is deactivated and the flow is null.
In order to build the function between ﬂow and head for the head-dependent well, we decided to place 11 drains in the same cell, one underneath the other. The topmost drain is located at elevation $0$ m, the spacing between the drains is $1$ m in the vertical direction and all drains have the same conductance equal to $0.010 m^2/d$. The resulting superposition of the 11 separate piecewise functions corresponds to a single function between the flux and the head in the cell. This may be treated as a single head-dependent well which removes water according to the relationship illustrated in Figure \ref{fig:testcase_nonlinear_setup} - Right with an approximate function found by interpolation:  
\begin{equation}
Q(u)=
    \begin{cases}
        0 & \text{if } u\leq -11\\
        - 5.0 \cdot 10^{-7} x^{5} + 2.0 \cdot 10^{-6} x^{4} + 2.4 \cdot 10^{-4} x^{3} + 2.9 \cdot 10^{-3}  x^{2} - 0.11 x - 0.66  & \text{if } u > -11 \\
    \end{cases}
        \label{eq:eq_nonlinear}
\end{equation}

The calculations are carried out with iMOD Python for a single stress period of 200 days. The hydraulic head is the output of the calculation, as shown in Figure \ref{fig:4testcases} (fourth row).

\end{description}



\section{Comparing neural network models} 
\label{comparison}

In this section, we compare the accuracy of vanilla DeepONet with two popular neural network models. The first model is an image-to-image regression approach that employs an encoder-decoder (ED)architecture \cite{eigen2014depth} and is constructed as a spatial downsampling pass, followed by a spatial upsampling pass. The second model considered is the Fourier neural operator (FNO) \cite{li2020fourier}, which utilizes the fast Fourier Transform to parameterize the integral kernel directly in the Fourier space.

The encoder network consists of multiple convolutional layers, followed by batch normalization, and ReLU activation function. The layer with the lowest dimension has $128$ feature maps.  The decoder network is similar to the encoder network but uses transposed convolutional layers instead of convolutional layers. Each convolutional layer uses a kernel of size $4\times4$ and a stride of $2\times2$ and zero padding. We follow the implementation of FNO as presented in \cite{li2020fourier} for the $2D$ Darcy Flow problem. 

The relative $L^2$ test error of the different network architectures is summarized in Table \ref{tab:model_comp_appendix}. Among all trained models, one can see that the DeepONet architecture yields the best predictive accuracy. Furthermore, compared to other the two neural operator methods which have to be trained using the points in the whole domain, DeepONet has the advantage that it can be trained on a few sparse observations. However, in order to carry out a fair comparison, we trained DeepONet considering the full field observation data. 

\begin{table}[H]
\begin{adjustbox}{max width=1.0\textwidth,center}
    \centering
    \begin{tabular}{|c|c|c|}
\hline
 \diagbox{Model}{Relative $L^2$ error}  &  Relative $L^2$ error for the Inverse problem & Relative $L^2$ error for the Nonlinear system  \\ \hline
  DeepONet        &    $2.23E-01 \pm 4.30E-02$ & $2.68E-02 \pm 7.53E-03$    \\ \hline
ED                    &         $3.70E-01 \pm 7.17E-02$ & $3.16E-02\pm 1.22E-02$           \\ \hline
FNO              &     $3.16E-01\pm 6.53E-02$ & $3.67E-02\pm 8.26E-03$                           \\ \hline
\end{tabular}

    \end{adjustbox}
        \caption{Benchmark on Inverse Problem and Nonlinear system: Mean and standard deviation of relative $L^2$ prediction errors of DeepONet, Encoder-Decoder, and Fourier Neural Operator.
        } 
        \label{tab:model_comp_appendix}
\end{table}

\section{SVD analysis of training data} \label{SVD}

The lower accuracy that the vanilla DeepONet shows when approximating the operator in the case in which the source term can appear at any location is traceable to a limitation of the model in dealing with symmetries such as translation, rotation, and stretching. Venturi \cite{Flex-DeepONet} describes DeepONet as a linear projection-based method, presenting the same inefficiency of singular value decomposition (SVD) when objects translate, rotate, or scale \cite{Brunton_book}. By showing that the trunk retrieves the modes of the projection, Venturi recommends employing SVD on the training data. This approach can help design the trunk net and its number of outputs during an exploratory phase. 

\begin{figure}[H]
    \centering
    \includegraphics[height=2in]{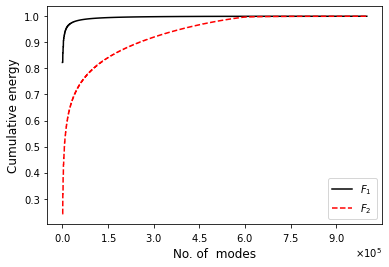}
    \caption{SVD on the output functions of the two data-set $F_{1}$ of experiment \textbf{E1} and $F_{2}$ of experiment \textbf{E2}: 1) the pumping well is placed in the same location of the domain across the whole training and testing datasets, 2) the pumping well can be placed at any location.}
    \label{fig:svd}
\end{figure}

Performing SVD on the output solutions gives the number of modes that are required to capture the variance or energy in the dataset. Each output solution of the labeled dataset, with dimension $32\times32$, is reshaped into a column vector; all the resulting vectors are then stacked horizontally as columns in the matrix \textbf{X}. Figure~\ref{fig:svd} shows the results of the SVD analysis on the matrix \textbf{X} of datasets of the first two test cases. When the pumping well is fixed at the center  of the domain, the energy contained in the first mode is equal to $85\%$. Its value decreases remarkably when the pumping well can be placed at any location: the first mode captures just $24\%$ of the energy of the original datasets and the singular values produced by SVD decay slowly. The results suggest that the SVD fails as it doesn't account for the translating nature of the data. We conclude that this is also the reason why vanilla DeepONet cannot accurately predict the solution for unseen well locations. 

\end{document}